\documentclass[a4paper,fleqn,usenatbib]{mnras}
\usepackage[T1]{fontenc}
\usepackage{ae,aecompl}
\usepackage{graphicx}	
\usepackage{amsmath}	
\usepackage{amssymb}	
\usepackage{amsbsy}
\usepackage{float}
\usepackage{color}

\usepackage{mathrsfs,bm}
\newcommand{\bcdot}{\ensuremath{%
  \mathchoice%
   {\mskip\thinmuskip\lower0.2ex\hbox{\scalebox{1.5}{$\cdot$}}\mskip\thinmuskip}}%
   {\mskip\thinmuskip\lower0.2ex\hbox{\scalebox{1.5}{$\cdot$}}\mskip\thinmuskip}%
   {\lower0.3ex\hbox{\scalebox{1.2}{$\cdot$}}}%
   {\lower0.3ex\hbox{\scalebox{1.2}{$\cdot$}}}%
}

\title[Faraday rotation maps of disk galaxies]{Faraday rotation maps of disk galaxies}

\author[R.~Pakmor et al.]  {R\"udiger~Pakmor$^{1,2}$\thanks{E-mail: rpakmor@mpa-garching.mpg.de}, Thomas Guillet$^{1,3}$, Christoph Pfrommer$^4$, \newauthor Facundo A. G\'omez$^{5,6}$, Robert J.~J. Grand$^{1,3}$, Federico Marinacci$^{7,8}$, \newauthor Christine M. Simpson$^1$, Volker Springel$^{1,3,2}$
  \vspace*{0.2cm}  \\
  $^1$Heidelberger Institut f\"{u}r Theoretische Studien,
  Schloss-Wolfsbrunnenweg 35, 69118 Heidelberg, Germany\\
  $^2$Max-Planck-Institut f\"{u}r Astrophysik, Karl-Schwarzschild-Str. 1, D-85748, Garching, Germany\\
  $^3$Zentrum f\"ur Astronomie der Universit\"at Heidelberg, ARI, M\"onchhofstrasse
  12-14, 69120 Heidelberg, Germany\\
  $^4${Leibniz-Institut f\"{u}r Astrophysik Potsdam (AIP), An der Sternwarte 16, 14482 Potsdam, Germany}\\
  $^5$Instituto de Investigaci{\'o}n Multidisciplinar en Ciencia yTecnolog{\'i}a, Universidad de La Serena, Ra{\'u}l Bitr{\'a}n 1305, La Serena, Chile\\
  $^6$Departamento de F{\'i}sica y Astronom{\'i}a, Universidad de LaSerena, Av. Juan Cisternas 1200 N, La Serena, Chile\\
  $^7$Kavli Institute for Astrophysics and Space Research, Massachusetts Institute of Technology, Cambridge, MA 02139, USA\\
  $^8$Harvard-Smithsonian Center for Astrophysics, 60 Garden Street, Cambridge, MA 02138, USA \\
}
  
\pubyear{2017}

\begin{document}

\label{firstpage}
\pagerange{\pageref{firstpage}--\pageref{lastpage}}

\maketitle

\begin{abstract}
Faraday rotation is one of the most widely used observables to infer the strength and configuration of the magnetic field in the ionised gas of the Milky Way and nearby spiral galaxies. Here we compute synthetic Faraday rotation maps at $z=0$ for a set of disk galaxies from the Auriga high-resolution cosmological simulations, for different observer positions within and outside the galaxy. We find that the strength of the Faraday rotation of our simulated galaxies for a hypothetic observer at the solar circle is broadly consistent with the Faraday rotation seen for the Milky Way. The same holds for an observer outside the galaxy and the observed signal of the nearby spiral galaxy M51. However, we also find that the structure and angular power spectra of the synthetic all-sky Faraday rotation maps vary strongly with azimuthal position along the solar circle. We argue that this variation is a result of the structure of the magnetic field of the galaxy that is dominated by an azimuthal magnetic field ordered scales of several kpc, but has radial and vertical magnetic field components that are only ordered on scales of 1-2 kpc. Because the magnetic field strength decreases exponentially with height above the disk, the Faraday rotation for an observer at the solar circle is dominated by the local environment. This represents a severe obstacle for attempts to reconstruct the global magnetic field of the Milky Way from Faraday rotation maps alone without including additional observables.
\end{abstract}


\section{Introduction}

The magnetic field in the Milky Way and in nearby disk galaxies of similar mass is in equipartition with thermal, turbulent, and cosmic ray energy densities \citep{Boulares1990,Beck1996,Fletcher2010}. Therefore, understanding the magnetic field of disk galaxies is necessary to understand their dynamical evolution as well as for modelling anisotropic transport processes along magnetic field lines.

\begin{figure*}
  \centering
  \includegraphics[width=0.99\textwidth]{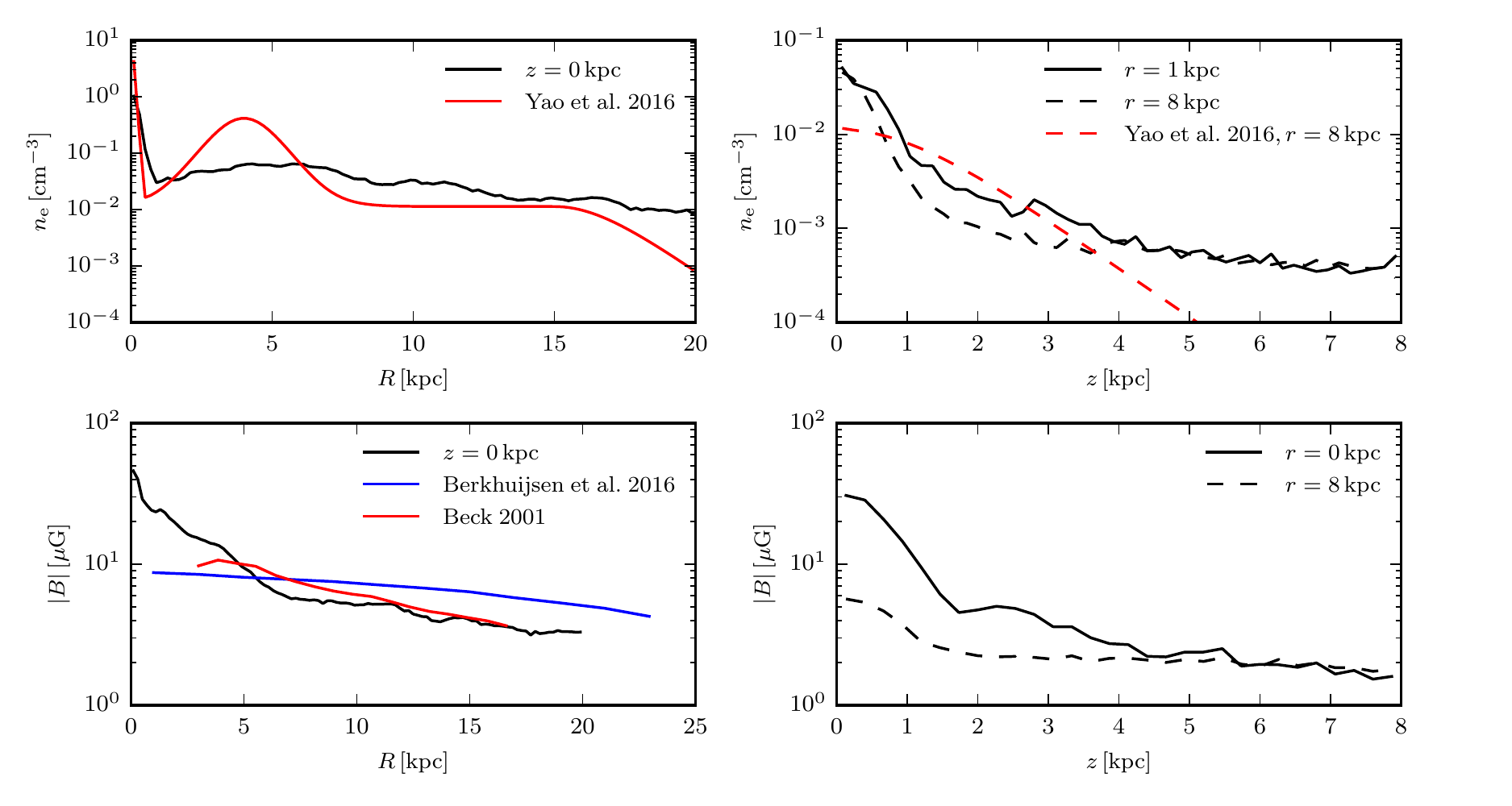}
  \caption{Radial (first column) and vertical (second column) profiles of the thermal electron density (top row) and the magnetic field strength (bottom tow) of Au-6. The profiles
  are computed as volume weighted averages of the electron density and the volume weighted root mean square of the magnetic field strength, thereby conserving the total electron number and magnetic energy. Radial profiles are calculated within a height of $0.5\,\mathrm{kpc}$ above and below the midplane of the disk. Vertical profiles are
  calculated for $0.5\,\mathrm{kpc} < r < 1.5\,\mathrm{kpc}$ (straight lines) and $7.5\,\mathrm{kpc} < r < 8.5\,\mathrm{kpc}$ (dashed lines), respectively. The red lines show the thermal electron density inferred for the Milky Way \citep{Yao2017} in the upper row. Of this model we include only the terms for the thin disk, the thick disk, and the Galactic center. The lower left panel also shows the radial magnetic field strength profile inferred for M101 (blue) \citep{Berkhuijsen2016} and the Milky Way (red)} \citep{Beck2001}.
  \label{fig:profiles}
\end{figure*}

The magnetic field of galaxies is very hard to observe directly. Instead, a number of indirect tracers of the magnetic field are employed to infer its properties in galaxies \citep[for an overview see][]{Beck2016,Han2017FaradayReview}. One widely used observational tracer is Faraday rotation of polarized radio continuum sources at frequencies of a few GHz where Faraday depolarisation is weak. It carries information about magnetic field strength as well as the structure of the magnetic field. Together with other tracers, it has been used to build and test models of the global magnetic field of the Milky Way \citep[see e.g.][]{Mao2010,VanEck2011,Pshirkov2011,Jansson2012a}.  Often the magnetic field is deemed to be dominated by either a dipole or a quadrupole field on large scales as expected from idealised galactic dynamo models \citep{Shukurov2006}. Nevertheless many observations point to a more complex structure \citep{Men2008,Mao2010,PlanckXLII}. 

One major advantage of the Faraday rotation over other tracers such as polarised and unpolarised synchrotron emission, is that it does not depend on the properties of the population of cosmic ray electrons in the observed galaxy, which is typically also poorly known. Instead, it depends on the thermal electron density which can be measured independently from tracers of the ionised medium. However, interpretation of observed Faraday rotation can be non-trivial as well \citep{Oppermann2012}, so numerical simulations of galaxies including magnetic fields are needed to better understand the observed data.

Until recently, the evolution of magnetic fields could only be simulated in idealised simulations of isolated disk galaxies \citep[see e.g.][]{Wang2009,Hanasz2009,Kulpa2011,Rieder2016,Butsky2017} ignoring their complicated evolutionary history and cosmological environment. Recently, however, high-resolution cosmological simulations of disk galaxies that follow the evolution of magnetic fields from the formation of the galaxies to $z=0$ have become feasible \citep{Pakmor2014,Pakmor2017,Rieder2017}.

Here we present for the first time synthetic Faraday rotation maps for polarized background sources of a fully cosmological high-resolution simulation of a Milky Way-like disk galaxy. In Sec.~\ref{sec:simulations}, we summarise the main properties of the simulation. In Sec.~\ref{sec:faraday}, we show Faraday rotation maps of the simulated galaxy as seen by different observers and compare them to observations. In Sec.~\ref{sec:discussion}, we discuss the physical origin of the Faraday rotation and its implications. Finally, we conclude in Sec.~\ref{sec:conclusions}.

\begin{figure*}
  \includegraphics[width=\textwidth]{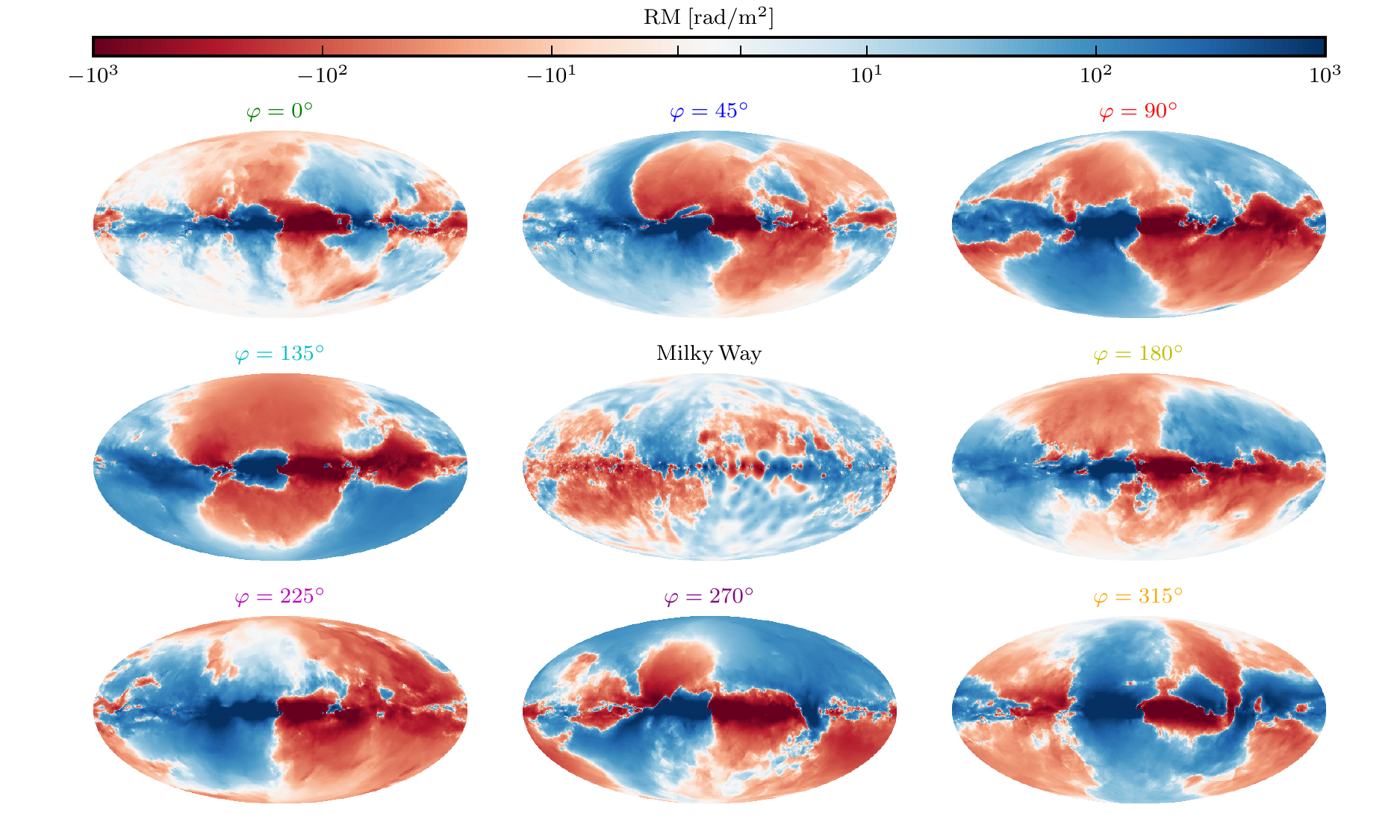}
  \caption{The reconstructed Faraday rotation map of the Milky Way \citep{Oppermann2015} (central map) and all-sky Faraday rotation maps of Au-6 for an observer at different positions in the midplane of the disk at a radius of $8\,\mathrm{kpc}$. The azimuthal positions $\varphi$ on this circle are separated by $45$ degrees. The maps are calculated with a resolution of $1024$ and $512$ bins for latitude and longitude, respectively and integrated to a depth of $20\,\mathrm{kpc}$. Contributions from larger distances are negligible. The color scale is logarithmic for $\left|RM\right| > 10\,\mathrm{rad/m^2}$ and linear for $\left|RM\right| < 10\,\mathrm{rad/m^2}$. }
  \label{fig:faraday}
\end{figure*}

\begin{figure}
  \includegraphics[width=\linewidth]{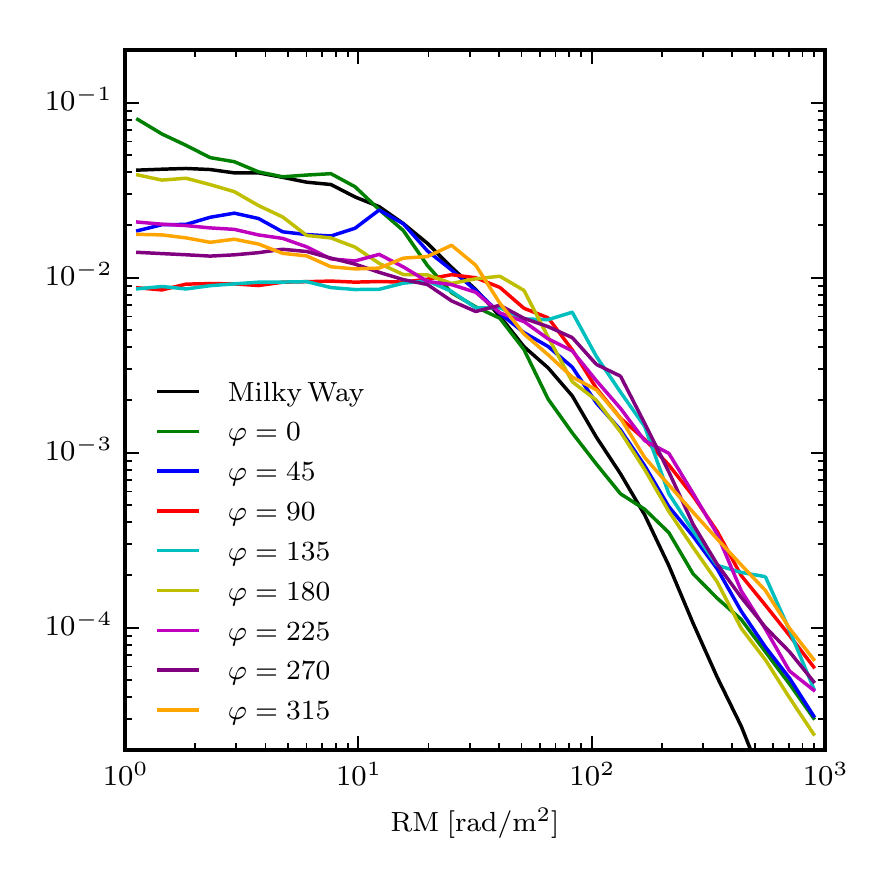}
  \caption{Solid angle weighted normalised histograms of the Faraday rotation maps shown in Fig.~\ref{fig:faraday}.}
  \label{fig:histogram}
\end{figure}

\begin{figure}
  \includegraphics[width=\linewidth]{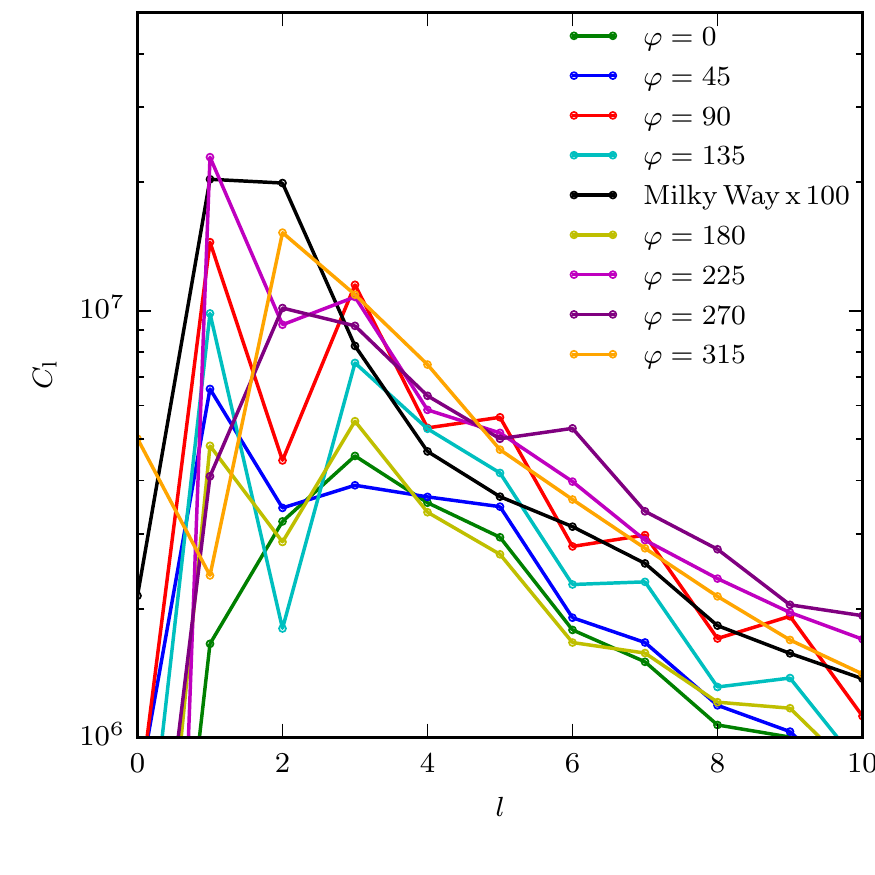}
  \caption{Angular power spectra for each of the all-sky Faraday rotation maps shown in Fig.~\ref{fig:faraday} to a multipole moment of $l=10$, coloured according to the labels of each map. The black line, which shows the power spectrum of the Faraday rotation map of the Milky Way, has been multiplied by a factor of $100$.}
  \label{fig:power}
\end{figure}

\section{Simulations}

\label{sec:simulations}

We analyse galaxies of the Auriga project \citep{Auriga} that comprises high-resolution cosmological zoom-in simulations of Milky Way-like halos. The simulations are carried out with the moving-mesh magnetohydrodynamics (MHD) code \textsc{arepo} \citep{Arepo}. They include dark matter, stars, and black holes that are all modelled with collisionless particles. In addition, gas is evolved on an unstructured Voronoi mesh using a second-order finite volume scheme \citep{Pakmor2016}. Gas cells move with a velocity close to their gas velocity to achieve a quasi Lagrangian description of hydrodynamics. Explicit refinement and de-refinement is employed to keep the masses of all cells in the high resolution region within a factor of two of the target mass resolution (here 3--5$\times 10^3\,\mathrm{M_\odot}$, the precise value of which depends on the halo).

The Auriga simulations use a comprehensive model for baryonic processes relevant for the formation and evolution of galaxies \citep{Marinacci2014,Auriga}. The model employs a sub-grid treatment of star formation and the interstellar medium \citep{Springel2003}. It includes atomic and metal line cooling and a spatially uniform UV background, and self-consistently evolves stellar populations leading to the injection of metals via AGB stars, core collapse supernovae, and type Ia supernovae \citep{Vogelsberger2013}. Supernova feedback is taken into account by a phenomenological model that launches galactic winds \citep{Auriga}. Moreover, we model accretion on and feedback from supermassive black holes \citep{BH_paper}.

Magnetic fields are evolved in the approximation of ideal MHD using cell-centered magnetic fields \citep{Pakmor2011}. To enforce the divergence constraint of the magnetic field we employ the Powell scheme \citep{Powell1999,Pakmor2013}. Each simulation is initialised with a homogeneous seed field at $z=127$ with a physical field strength of $2 \times 10^{-4} \mu \rm{G}$. As shown in \citet{Pakmor2017}, the magnetic field in the galaxy is exponentially amplified to about $10\%$ of its equipartition value between $z=7$ and $z=3$ by a turbulent dynamo that is driven by accretion of gas on the galaxy \citep[see also][]{Pakmor2014,Marinacci2015}. After the formation of a disk around $z=1$ a combination of differential rotation and vertical flows in the disk then further amplifies the magnetic field to reach equipartition between $z=0.5$ to $z=0$. Its present day strength is consistent with the observed field strength of nearby disk galaxies. The magnetic field becomes ordered during its second phase of evolution after the disk has formed. It changes from the turbulent magnetic field left behind by the turbulent dynamo to a configuration that is dominated by a large-scale azimuthal field \citep{Pakmor2017}.

We include all high resolution halos in our analysis, but we focus in particular on halo Au-6 from the Auriga suite because it hosts a galaxy that closely resembles the Milky Way \citep{Auriga}. At $z=0$ it has a halo mass of $1.0\times10^{12}\,\mathrm{M_\odot}$ and a stellar mass of $6\times10^{10}\,\mathrm{M_\odot}$. It is disk-dominated with a disk/total ratio of $0.78$ and radial scale length of $3.2\,\mathrm{kpc}$. Note that the other high resolution halos are in many respects also very similar to the Milky Way and span a range of stellar and halo masses and disk properties \citep{Auriga}.

\begin{figure*}
  \includegraphics[width=0.85\linewidth]{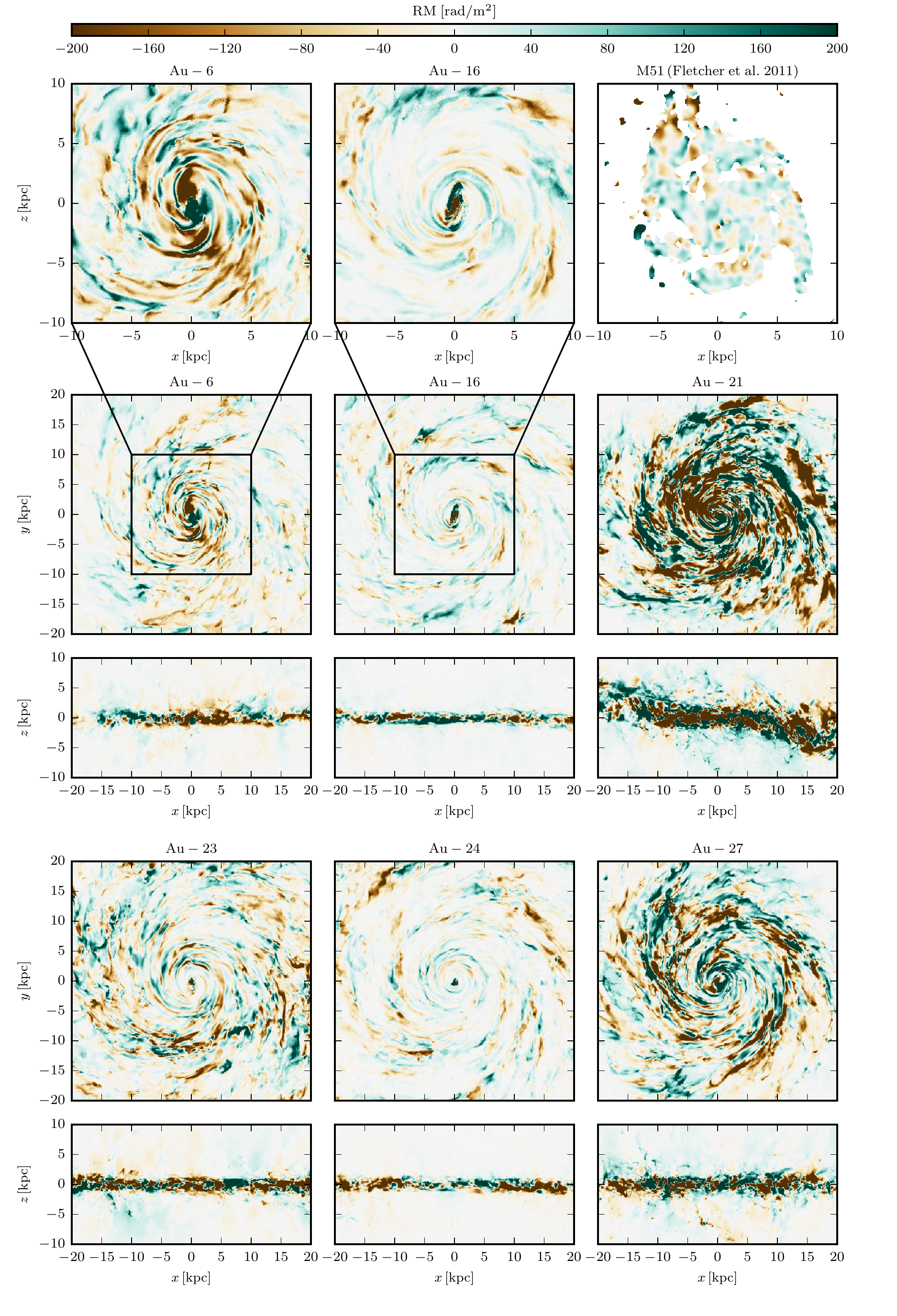}
  \caption{Faraday rotation maps as seen by an external observer. The top panel shows zoom-ins of face-on Faraday rotation maps of halos Au-6 and Au-16 and the observed Faraday rotation map of M51 for wavelengths between 3cm and 6cm \citep{Fletcher2011} assuming a distance to M51 of $7.97\,\mathrm{Mpc}$ \citep{Bose2014}. Middle and bottom panel show face-on and edge-on projected Faraday rotation maps of all high-resolution simulations of the Auriga project.}
  \label{fig:faraday_extern}
\end{figure*}

\section{Synthetic Faraday rotation maps}

\label{sec:faraday}

The Faraday rotation measure (RM) is given by the line of sight integral of the magnetic field component along the line of sight times the local thermal electron density, i.e.
\begin{equation}
RM = \frac{e^3}{2\pi m_\mathrm{e}^2 c^3} \int n_\mathrm{e}\left( l \right) \vec{B}\left( l \right) \cdot \mathrm{d}\vec{l},
\end{equation}
where $e$ and $m_\mathrm{e}$ are the charge and the mass of an electron, $c$ is the speed of light, $n_\mathrm{e}$ is the electron density, and $B_{||} = \vec{B} \cdot \vec{l}$ is the magnetic field component along the line of sight $l$. The line of sight integral starts at the position of the observer and ends at the position of a background source whose polarisation is used to measure Faraday rotation via a rotation in the direction of the polarisation. We do not attempt to generate true synthetic observables that would include depolarisation effects such as turbulent magnetic fields, RM dispersion and RM gradients along the line of sight or within the telescope beam \citep[e.g.][]{Beck2016}), but for this work we use the theoretical Faraday rotation as defined above. Note, however, that depolarisation can potentially have a significant effect on the observed signal \citep{Nixon2018}. Moreover, we only model the signal for polarised extragalactic radio sources and do not include sources in the galaxy itself (e.g. pulsars). This way we avoid potential lines of sight that stop in the galaxy and therefore only contribute parts of the signal compared to an extragalactic source.

The rotation measure depends only on the thermal electron density and the magnetic field; both are readily available quantities in the simulation. We show their radial and vertical profiles for our simulated galaxy at $z=0$ in Fig.~\ref{fig:profiles}. For gas that is not star-forming, we can directly use the thermal electron density as calculated in the cooling module of the code. For star-forming gas, we need to take into account the subgrid model adopted by the code. This model implicitly assumes an unresolved multi-phase interstellar medium made up of a volume-filling warm phase and a number of dense cold clumps that contain most of the mass of a cell \citep{Springel2003}. To calculate the contribution to the rotation measure integral of star-forming gas we only use the gas that our subgrid model assigns to the volume-filling warm phase. We calculate the mass fraction of the warm phase as described in \citet{Springel2003} and assume that it is fully ionised.

As shown in Fig.~\ref{fig:profiles}, the radial electron density profile is in good agreement with recent models of the electron density of the Milky Way \citep{Yao2017}. The main differences are that our simulated galaxy does not feature as thin a disk component as the Milky Way, which is responsible for the bump in the observed profile at $r\approx 4\,\mathrm{kpc}$. The vertical electron density profile of our galaxy is consistent within a few kpc, but decreases less steeply for $\left|z\right| > 5\,\mathrm{kpc}$ than in the Milky Way. The magnetic field strength profile is consistent with the inferred magnetic fields of nearby galaxies \citep{Pakmor2017}. It is important to note that the rotation measure depends on the magnetic field component resulting from the projection of the full magnetic field vector on the line of sight and therefore can be positive or negative. Thus, not only does the strength of the magnetic field influence the Faraday rotation, but it is also sensitive to the detailed structure of the magnetic field. 

\begin{figure}
  \includegraphics[width=\linewidth]{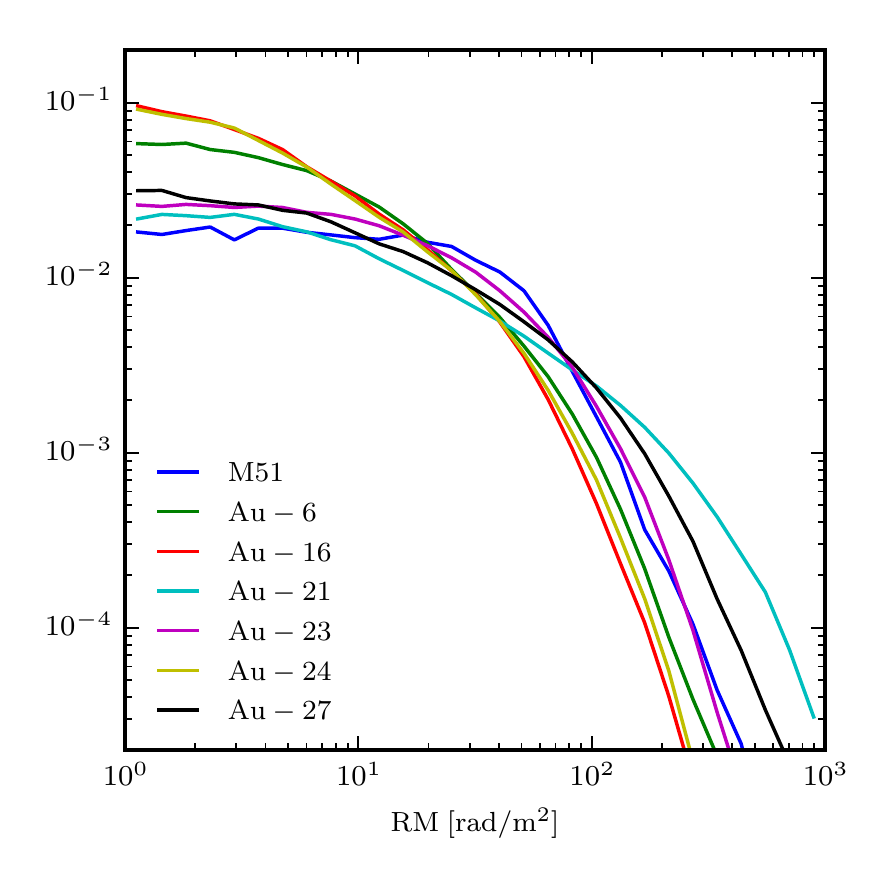}
  \caption{Area weighted normalised histograms of the Faraday rotation maps for an external observer viewing the galaxies face-on as shown in Fig.~\ref{fig:faraday_extern}.}
  \label{fig:histogram_extern}
\end{figure}

\begin{figure}
  \includegraphics[width=\linewidth]{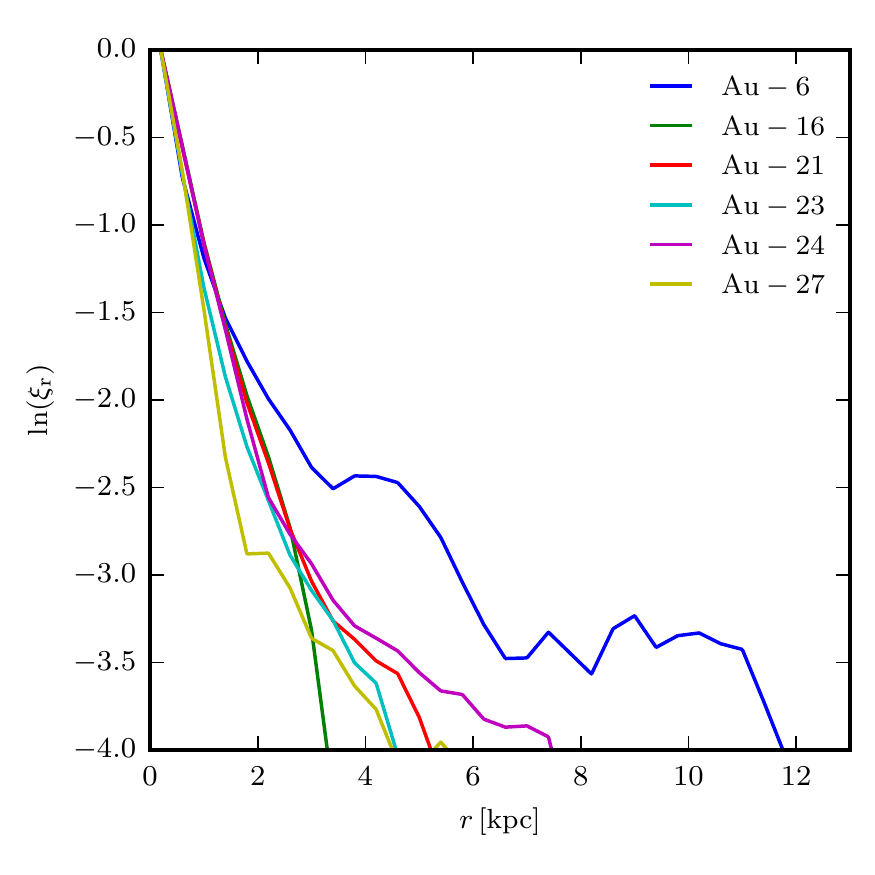}
  \caption{Real-space two-point correlation function for the face-on Faraday rotation maps of the simulated galaxies shown in Fig.~\ref{fig:faraday_extern}. Only pixels in the range $3\,\mathrm{kpc} < r < 20\,\mathrm{kpc}$ are included to avoid being dominated by the center of the galaxy and to concentrate the signal on the part that is also seen in the all-sky maps. The correlation functions are normalised to one at the smallest radial bin.}
  \label{fig:corr}
\end{figure}

\begin{figure*}
  \includegraphics[width=\textwidth]{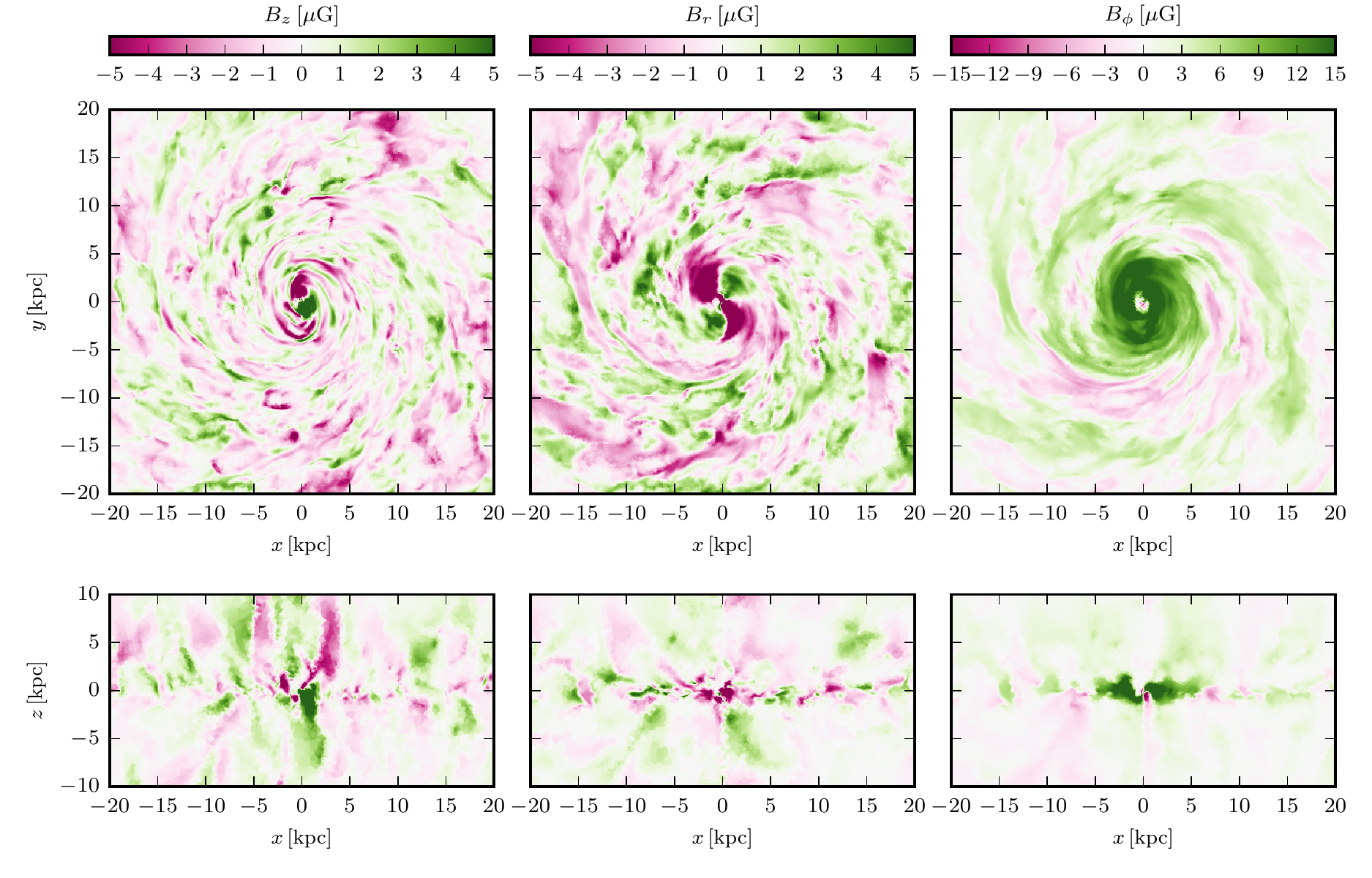}
  \caption{Thin projections (with a depth of $1\,\mathrm{kpc}$) of the magnetic field components in cylindrical coordinates centered on the midplane (top row) and perpendicular to the disk (bottom row) for Au-6. The columns show the vertical magnetic field strength (left column), the radial magnetic field strength (middle column), and the azimuthal magnetic field strength (right column), respectively.}
  \label{fig:bfld}
\end{figure*}

In Fig.~\ref{fig:faraday}, we show synthetic, all-sky Faraday rotation maps for observers at different positions in the midplane of the disk along a circle at the solar radius of $8\,\mathrm{kpc}$ \citep{MWCenter} and compare them to the reconstructed Faraday map of the Milky Way \citep{Oppermann2012,Oppermann2015}. The strength of the synthetic rotation measure varies from $\left| RM \right| \gtrsim 1000\,\mathrm{rad/m^2}$ in the center of the disk to $\left| RM \right| \lesssim 50\,\mathrm{rad/m^2}$ above and below the disk and is broadly consistent with the observed values for the Milky Way with a slightly larger amplitude. This can be seen quantitatively in Fig.~\ref{fig:histogram} which shows a histogram of the rotation measure values for all Faraday rotation maps of Fig.~\ref{fig:faraday}. The synthetic maps have a similar distribution as the observed map of the Milky Way, but exhibit more high RM values. There is considerable scatter between the synthetic maps for different observers in the same galaxy. Unlike in the observed map of the Milky Way, Fig.~\ref{fig:faraday} indicates a trend of decreasing size of the rotation measure in the plane of the disk with distance from the center in the synthetic maps. We also lack the large variations on the smallest scales seen in observations. Most likely this is due to limitations of our physical model that suppress the small-scale turbulent magnetic field on scales smaller than $100\,\mathrm{pc}$. In particular, we do not resolve the multi-phase structure of the ISM but employ a subgrid model for the ISM \citep{Springel2003}. On the resolved scales this effective model then produces a smooth ISM. Moreover, in this model we do not add the energy input from supernovae explicitly. Instead it enters into the effective ISM and galactic wind model. Thus, our ISM is less turbulent than expected for a fully resolved multi-phase ISM. Moreover, the turbulent magnetic field can be increased by local amplification from SN remnants\citep{Butsky2017}. In principle noise in the data may also contribute to the difference, but is likely subdominant here.

There are obvious qualitative differences between the different synthetic Faraday rotation maps of our simulated galaxy. Only the central part of the disk consistently shows a strong positive value on the left and a strong negative value on the right of the center, as expected for a dominant coherent azimuthal magnetic field. In contrast, the structure of the Faraday rotation measure in the outer parts of the disk and even more so above and below the disk is completely different from map to map. This statement also holds for the angular power spectra obtained from the synthetic rotation measure maps that are shown in Fig.~\ref{fig:power}. As shown in this figure, the power spectra of the synthetic maps look similar to the power spectrum of the Faraday map of the Milky Way, although the former have an amplitude that is about a factor of $100$ larger. The main source of this discrepancy in the amplitude is the stronger signal in the plane of the disk in the synthetic maps that indicates that the large-scale component of the azimuthal magnetic field in our simulation is stronger and more coherent than for the Milky Way. However, also depolarisation of the observed signal may contribute here.

The power spectra of the synthetic maps can be dominated by a dipole (e.g. $\varphi=225^{\circ}$, purple, bottom left panel in Fig.~\ref{fig:faraday}), by a quadrupole (e.g. $\varphi=315^{\circ}$, orange, bottom right in Fig.~\ref{fig:faraday}), or by higher order multipoles (e.g. $\varphi=0^{\circ}$, green, top left panel  in Fig.~\ref{fig:faraday}). This strongly suggests that the synthetic Faraday rotation of our galaxy is dominated by the local signal, i.e. contributions from the region close to the observer. We will discuss this idea in more detail in Sec.~\ref{sec:discussion}. It is also consistent with the finding that there is no large-scale coherent vertical magnetic field at the position of the Sun in the Milky Way as inferred from the Faraday rotation measured towards the galactic poles \citep{Mao2010}. 

Synthetic Faraday rotation maps for an external observer looking at the galaxy face-on or edge-on are shown in Fig.~\ref{fig:faraday_extern} for all six high-resolution Auriga halos and M51 \citep{Fletcher2011} and their histograms are shown in Fig.~\ref{fig:histogram_extern}. The typical amplitude of the Faraday rotation in the synthetic maps varies between the galaxies. For Au-16 and Au-24 that show the weakest signal the amplitude of the Faraday rotation maps is very similar to the observed map of M51. For Au-6 and Au-23 the amplitude is slightly larger and for Au-21 and Au-27 it is significantly larger than for M51. The histogram of RM values of M51 lies within the range spanned by our galaxies. Note however, that M51 is a few times less massive than all our simulated galaxies and that the map shown in Fig.~\ref{fig:histogram_extern} is based on the polarisation signal originating from the diffuse emission of the galaxy. Nevertheless, at the wavelength of 3cm/6cm used the depolarisation effects are small and the measurement should be comparable to our synthetic map of background sources. In all synthetic maps the structure of the signal is qualitatively consistent with the observed Faraday rotation map of M51, in particular they all show typical variation of the sign of the RM on kpc scales, although the spiral arms are emphasised a bit more strongly in the synthetic maps. For Au-6 there is a visible decrease of the strength of the Faraday rotation measure with radius in the disk. This is similar to the all-sky Faraday rotation maps discussed before for an observer at the solar circle. This radial decrease does not seem to be present in the observed map of M51 or in the other galaxies and probably depends on the detailed structure of the bar and bulge of the galaxies.

This typical scale for the synthetic Faraday maps can be seen quantitatively in the two-point correlation functions of the face-on Faraday rotation maps presented in Fig.~\ref{fig:corr}. All maps show a very similar correlation function with a correlation length of about $1\,\mathrm{kpc}$, where we define the correlation length $\xi_e$ as the distance at which the normalised correlation function falls below $1/e$. Therefore, the Faraday rotation of the galaxies viewed face-on is essentially uncorrelated already for distances of a few kpc. Note that for a perfect face-on view of the galaxy, the Faraday rotation maps only depend on the vertical component of the magnetic field. Therefore, in this case the correlation function of the Faraday rotation is closely related to the correlation function of the vertical magnetic field. In contrast, all-sky maps for observers in the galaxy always depend on all three components of the galactic magnetic field.

\begin{figure}
  \includegraphics[width=\linewidth]{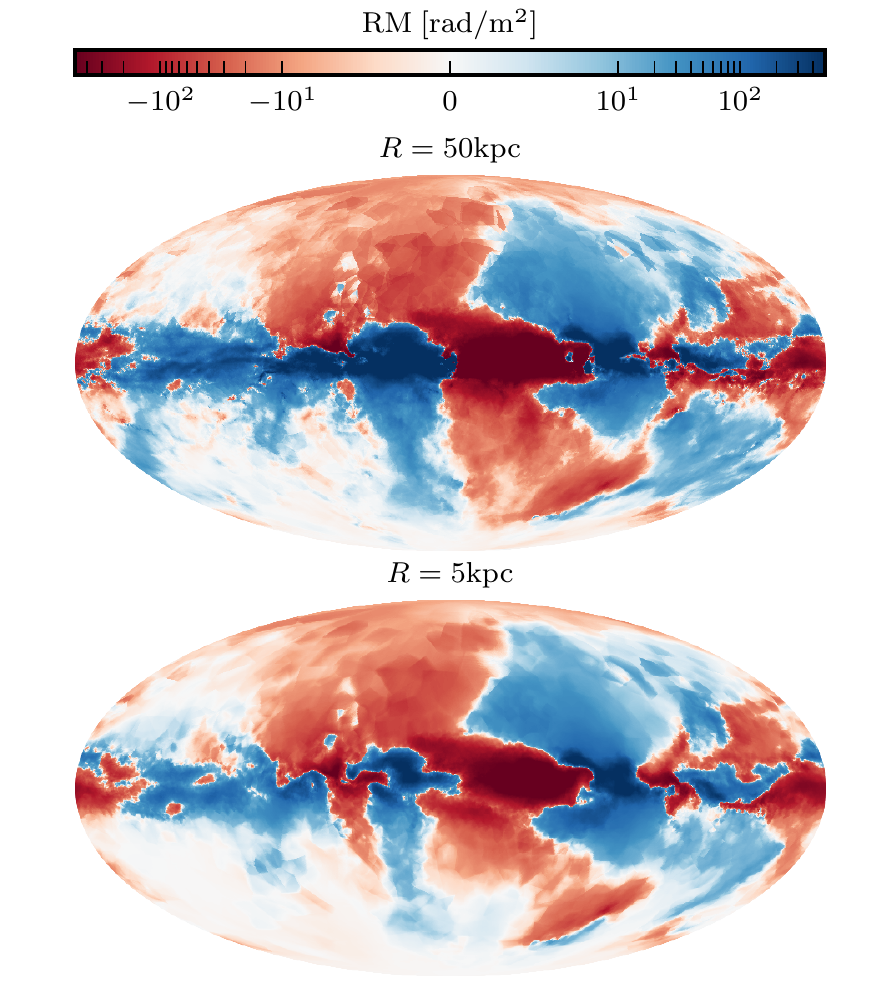}
  \caption{All-sky Faraday rotation maps of Au-6 for an observer in the midplane of the disk at a radius of $8\,\mathrm{kpc}$ at an angle of $\varphi=0$ degrees (see Fig.~\ref{fig:faraday}). The top panel shows the Faraday rotation map integrated to a distance of $50\,\mathrm{kpc}$; the map shown in the lower panel is only integrated to a distance of $5\,\mathrm{kpc}$.}
  \label{fig:close}
\end{figure}

\begin{figure}
  \includegraphics[width=\linewidth]{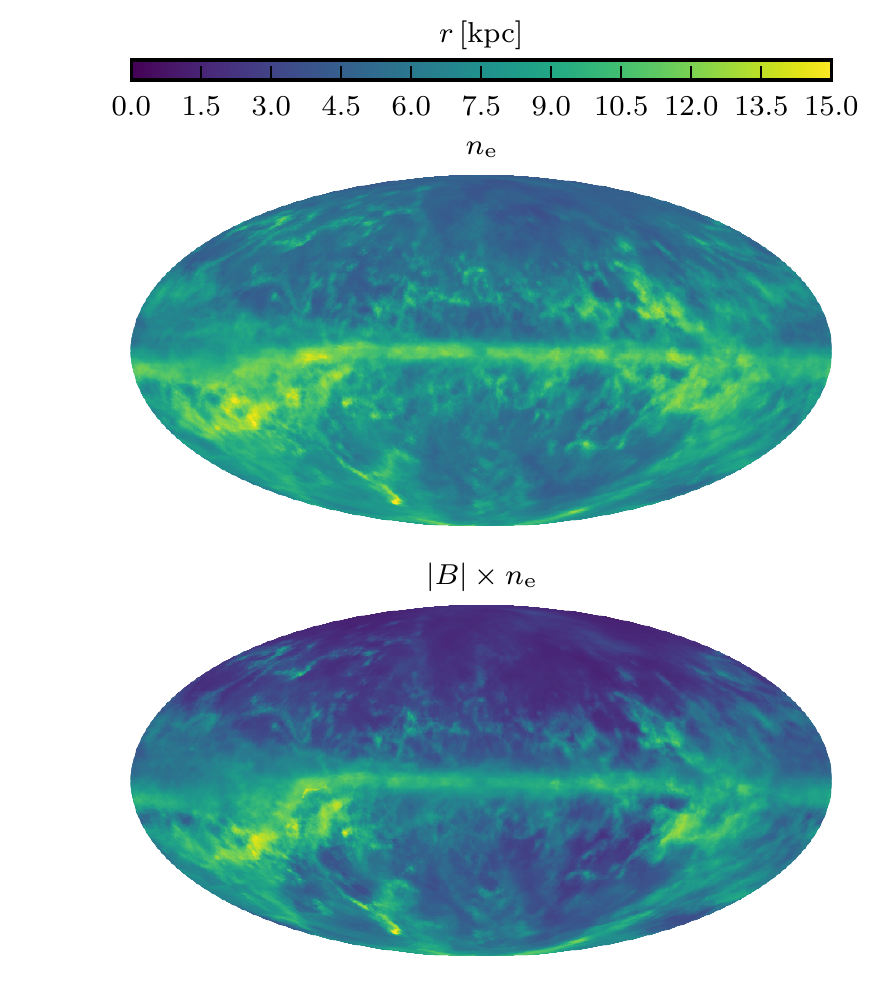}
  \caption{Average distance of contributions to the all-sky line of sight integral of the thermal electron density as defined in equation~\ref{eqn:ne} (top panel) and of the product of magnetic field strength and thermal electron density for Au-6 as defined in equation~\ref{eqn:bne} for an observer in the midplane of the disk at a radius of $8\,\mathrm{kpc}$ at an angle of $\varphi=0$ degrees.}
  \label{fig:distance}
\end{figure}

\section{Discussion}

\label{sec:discussion}

To understand the synthetic Faraday rotation we need to look at the structure of the magnetic field in the galaxy. The magnetic field components in cylindrical coordinates are shown in Fig.~\ref{fig:bfld}. The magnetic field is dominated by its azimuthal component that is coherently oriented parallel to the velocity field for $r < 5\,\mathrm{kpc}$. At larger radii it still shows dominant large-scale ordering, but exhibits magnetic spiral arms with reversed orientation. Typically there are two or three reversals at a given azimuthal angle compared to one confirmed reversal in the Milky Way \citep{VanEck2011}. The orientation of the azimuthal magnetic field component is approximately constant with height at $r < 5\,\mathrm{kpc}$. At larger radii it is roughly constant above and below the disk, but can flip in the midplane of the disk.

The vertical and radial components of the magnetic field in the midplane of the disk are weaker in strength by a factor of a few than the azimuthal magnetic field and are dominated by structures on smaller scales than the coherent disk-scale structures of the azimuthal magnetic field. In particular, the sign of the vertical and radial magnetic field changes on scales of about a few kpc. Similar to the azimuthal field component, the vertical and radial components are coherent over several kpc in the vertical direction, and their sign sometimes flips at the midplane of the disk and sometimes does not.

Importantly, even though its azimuthal component is ordered on large scales, the magnetic field of the galaxy as a whole is not dominated by a large-scale dipole or quadrupole field as expected from idealised models of a galactic mean-field dynamo \citep[see e.g.][]{Shukurov2006,Gressel2013,Chamandy2016} or some idealised simulations of isolated disk galaxies \citep[see e.g.][]{Kulpa2011}. Rather, the small-scale structure in the vertical and radial magnetic field provides the first step to understand the variations in the Faraday rotation maps with position in the galaxy.

As a second step, we test the hypothesis that the Faraday rotation in the all-sky maps is dominated by contributions from the local environment around the observer. To this end, we compare the full signal integrated out to $50\,\mathrm{kpc}$ with the signal integrated out only to $5\,\mathrm{kpc}$ from an observer at the solar circle in Fig.~\ref{fig:close}. Both maps resemble each other closely, in particular the large-scale structures of the full map that includes all contributions out to $50\,\mathrm{kpc}$ are already clearly present in the map that is integrated only to a distance of $5\,\mathrm{kpc}$ from the observer. This directly shows that it is the material within $5\,\mathrm{kpc}$ of the observer that dominates the Faraday rotation in our synthetic all-sky maps. 

In addition, we look at the average distance for contributions to the integrated signal for a very large integration distance. We cannot directly do this for the Faraday rotation measure because of possible cancellations, i.e. contributions with different sign can cancel each other. Nevertheless, we can still get a general idea from looking at the average contribution to an integral of the electron density only and of the product of electron density and strength of the magnetic field. To this end, we show in Fig.~\ref{fig:distance} their average distance-weighted contributions defined as

\begin{equation}
\displaystyle
r_{n_\mathrm{e}} = \frac{ \int n_\mathrm{e} \left( l \right) \times l\, \mathrm{d}l }{ \int n_\mathrm{e} \left( l \right)\, \mathrm{d}l }
\label{eqn:ne}
\end{equation}

and 

\begin{equation}
\displaystyle
r_{\left|B\right|*n_\mathrm{e}} = \frac{ \int \left|B\right| \times n_\mathrm{e} \left( l \right) \times l\, \mathrm{d}l }{ \int \left|B\right| \times n_\mathrm{e} \left( l \right)\, \mathrm{d}l }.
\label{eqn:bne}
\end{equation}

The maps demonstrate that the average radius of the contributions to the electron density integral is $r \gtrsim 10\,\mathrm{kpc}$ in the plane of the disk, but only about $r \approx 5\,\mathrm{kpc}$ for most of the areas above and below the disk. For the product of magnetic field strength and electron densities, the characteristic radius of the contributions above and below the plane of the disk drops to only a few kpc. This can be understood from the profiles in Fig.~\ref{fig:profiles} that show that the magnetic field strength and the thermal electron density drop quickly for larger radii and heights. Of course, this result would only be directly representative for the actual Faraday rotation measure for a completely ordered magnetic field without any field reversals. However, it still is an indication that the contributions of the local environment (within a few kpc of the observer) to the line of sight integral of the Faraday rotation measure dominate the signal, consistently with our direct check in Fig.~\ref{fig:close}.

Together with the variations of the directions of the radial and vertical magnetic field components on similar scales, this naturally explains the qualitative variations of the Faraday rotation measure maps for different observer positions in our simulated galaxy shown in Fig.~\ref{fig:faraday}. It also is consistent with the lack of variation in the synthetic maps towards the center of the galaxy, where the magnetic field is coherent.
    
The magnetic field structure of our simulated galaxy also indicates that there may be significant galactic variance in the magnetic field on larger scales, in addition to what is introduced by the turbulent component of the magnetic field \citep{PlanckXLII}.

\section{Conclusions}

\label{sec:conclusions}

We have shown Faraday rotation maps of simulated galaxies from full cosmological simulations whose magnetic field strength and thermal electron density are consistent with the Milky Way. The strength of the signal is consistent with observations for both all-sky maps from an observer at the solar circle as well as from an observer outside the galaxy. We find, however, that there are large variations in the realisation of the Faraday rotation maps when the position of the observer is varied at the same distance to the center of the galaxy. 

The main reason for the variations in the Faraday maps are variations in the radial and vertical components of the galactic magnetic field on scales of a few kpc and the fact that the Faraday rotation is dominated by contributions from the local environment out to a distance of a few kpc for most directions away from the galactic disk. Such a magnetic field structure disagrees with simple models of a galactic mean-field dynamo that predict the magnetic field of the galaxy to be dominated by a global dipole or quadrupole.

There are many potential reasons for this discrepancy. In principle, approximations in our baryonic physics model could influence the magnetic field structure of the magnetic field at late times. In particular the effective model for the interstellar medium and the phenomenological model for galactic winds could potentially alter their properties sufficiently to suppress or enhance a galactic dynamo that has been seen in some idealised simulations of isolated galaxies \citep[see e.g.][]{Kulpa2011}. In contrast, it is possible that the idealised conditions that allow the largest modes of the global magnetic field of a galaxy to dominate eventually may not be realised in galaxies in a realistic cosmological environment. In the future, cosmological simulations with more realistic physics models will be needed to analyse the formation of the late time large-scale magnetic field in more detail.

Moreover, our simple Faraday rotation maps and their analysis can only be a first step towards a much more sophisticated generation of full synthetic mock observations of galaxies from cosmological simulations. Then applying the same analysis as is done on observational data \citep[e.g.][]{Steiniger2018} will allow for a fair comparison between simulations and observations. It also remains to be seen if the magnetic field morphology of our simulated galaxy - that is significantly more complicated than what is expected from simple galactic dynamo models - is consistent with all observational data, as mean-field dynamo configurations have been claimed to be \citep{Braun2010}. Finally, detailed comparisons between cosmological simulations and observations will also be very important to understand forthcoming observations from new telescopes such as The Square Kilometre Array.
    
\section*{Acknowledgements}
  
We thank Ann Mao, Dandan Xu, Carlos Frenk, Simon White, and Stefan Rei{\ss}l for helpful discussion. We thank Andrew Fletcher and Niels Oppermann for providing the data of the Faraday rotation maps of M51 and the Milky Way.
This work has been supported by the European Research Council under ERC-StG grant EXAGAL-308037, ERC-CoG grant CRAGSMAN-646955, and by the Klaus Tschira Foundation. RP, RG and VS acknowledge support by the DFG Research Centre SFB-881 ``The Milky Way System" through project A1. TG and VS acknowledge support through subproject EXAMAG of the Priority Programme 1648 ``Software for Exascale Computing" of the German Science Foundation.  FAG acknowledges support from Fondecyt Regular 1181264, and funding from the Max Planck Society through a ``Partner Group" grant. This work used the Data Centric system at Durham University, operated by the Institute for Computational Cosmology on behalf of the STFC DiRAC HPC Facility ``www.dirac.ac.uk". This equipment was funded by BIS National E-infrastructure capital grant ST/K00042X/1, STFC capital grant ST/H008519/1, and STFC DiRAC Operations grant ST/K003267/1 and Durham University. DiRAC is part of the UK National E-Infrastructure.
    
\bibliographystyle{mnras}

\label{lastpage}

\end{document}